\documentclass{jaa}
\usepackage{graphicx}
\usepackage{amsmath}
\usepackage{xcolor}
\usepackage{natbib}
\usepackage{footmisc}
\usepackage{rotating,longtable}
\usepackage{lineno}
\usepackage{natbib}
\usepackage[T1]{fontenc}
\usepackage{ae,aecompl}

\begin{document}\sloppy

%%paper title
%%For line breaks \\ can be used within title
\title{A preliminary cosmological analysis of stellar population synthesis of galaxies released by LAMOST LRS DR11}

%%author names are separated by comma (,)
%%use \and before the last author name
%%use a * along with the number separated by comma
%% for the  author for correspondence
%%\textsuperscript{number} is used for affiliation
%%\affilOne, \affilTwo etc., upto \affilTwentyfive is possible
%%Please note the first letter after \affil is capitalised in the command
%%

%\author{Saurabh Sharma\textsuperscript{1,*}, et. al.}
\author{Y. H. Chen\textsuperscript{1,2,*}}

\affilOne{\textsuperscript{1}Institute of Astrophysics, Chuxiong Normal University, Chuxiong 675000, China\\}
\affilTwo{\textsuperscript{2}International Centre of Supernovae (ICESUN), Yunnan Key Laboratory, Kunming 650216, China\\}

%%escape two column mode for title, affiliation and abstract
%%by giving \twocolumn command as shown

\twocolumn[{

\maketitle

%%include \corres to print the corresponding author Email id
\corres{yanhuichen1987@126.com}

%%include \msinfo for
%%manuscript information such as
%%received, revised and accepted dates
%%
\msinfo{ }{ }

%%abstract
\begin{abstract}
The evolution of the universe together with the galaxies is one of the fundamental issues that we humans are most interested in. Both the observations of tidal streams from SDSS and the theory of $\Lambda$CDM support the hierarchical merging theory. The study of high redshift celestial bodies contributes to a more in-depth study of cosmology. The LAMOST low resolution search catalog DR11 v1.0 has released 11,939,296 spectra, including 11,581,542 stars, 275,302 galaxies, and 82,452 quasars, and so on. The data of 28,780 stellar population synthesis of galaxies and some high redshift quasars are used to do a preliminary statistical research. We selected the data with small errors for analysis and obtained some basic statistical conclusions. Older galaxies have relatively larger stellar velocity dispersions. The larger the metallicity, the greater the stellar velocity dispersion. These statistical results are reasonable and consistent with previous work. Because the stellar velocity dispersion is driven by the total mass of a galaxy at the first order and more massive galaxies have older ages and greater metallicities. The spectra of high redshift quasars show clear Gunn-Peterson trough and Lyman-$\alpha$ forest. The identified emission lines and high redshift celestial spectra released by LAMOST can be used for cosmological research.
\end{abstract}

%%insert keywords separated by 3 hyphens using \keywords{words}

}]
%%close the twocolumn escape here

%%include \doinum{number}for the DOI number in the header
%%include \volnum{number} for the volume number in the header
%%include \year{yyyy} for  year of publication in the header
%%include \pgrange{num--num} page range of article in the header
%%include \artcitid{num} for the article citation id
%%include \lp to print last page of the article
%%include \setcounter{page}{pagenum} for the exact starting page of the article

\doinum{12.3456/s78910-011-012-3}
\artcitid{\#\#\#\#}
\volnum{000}
\year{0000}
\pgrange{1--}
\setcounter{page}{1}
\lp{1}

	%%%%%%%%%%%%%%%%% BODY OF PAPER %%%%%%%%%%%%%%%%%%

\section{Introduction}

Modern cosmology believes that the universe was born in the Big Bang 13.8 billion years ago. Carrasco (2014) reported that the universe was dominated by radiation (redshift z $>$ 3000), matter (3000 $>$ z $>$ 0.5), and dark energy (z $<$ 0.5) respectively. In the early stages after the Big Bang, the universe was filled with ionized gases and rapidly expanded and cooled. Approximately 380 thousand years (Djorgovsk 2004) after the Big Bang, the universe cooled to $\sim$3,000\,K (z $\sim$ 1100, Djorgovski et al. 2001) and became neutral and opaque, marking the beginning of the Dark Ages. About a few hundred million years after the Big Bang, galaxies and quasars began to form, and then the reionization epoch started. The James Webb Space Telescope (JWST), launched in 2021, has the ability to study galaxy formation in the infrared band during the first billion years of cosmic history (Robertson 2022). Based on JWST, some galaxies with z = 9.79 (Roberts-Borsani et al. 2023), 10.38, 11.58, 12.63, and 13.20 (Curtis-Lake et al. 2023) were discovered. The metal-poor, young age, and small mass galaxy with z = 13.20 has pushed the start of the reionization epoch to 320 million years (Witze 2023) after the Big Bang. Studying the Gunn-Peterson trough, Fan et al. (2006) reported that the end of the reionization epoch is at z $\sim$ 6 to 8. Around one billion years after the Big Bang, the reionization completed and the universe becomes transparent again. After the epoch of reionization, galaxy evolution and star formation dominated the universe. Madau \& Dickinson (2014) derived that the star formation rate density had a peak approximately 3.5 billion years after the Big Bang, at z $\sim$ 1.9. The history of the universe is a very interesting process and one of the fundamental issues that we humans are most concerned about.

Eggen et al. (1962) studied the velocity vectors and moving orbits for 221 well observed dwarf stars in terms of the dynamics of a collapsing galaxy and first proposed a collapsed galaxy without rich sky survey observational data. On the contrary, based on the study of 177 red giants in 19 globular clusters, Searle \& Zinn (1978) reported that the metal abundances of these outer clusters were independent of galactocentric distance, and proposed an accretion and mergence model. The study of Sagittarius tidal stream based on the Sloan Digital Sky Survey (SDSS) directly supports the hierarchical merging theory (Ibata et al. 2001, Martínez-Delgado et al. 2001, Li et al. 2019). In addition, the $\Lambda$ cold dark matter ($\Lambda$CDM) model predicts more than hundreds of merging events during the formation of the Milky Way (Stierwalt et al. 2017, Xing et al. 2019, Enomoto et al. 2023). Therefore, galaxies are believed to form in a "bottom up" manner (Mori \& Umemura 2006). The formation and evolution of galaxies are also one of the fundamental issues that interest us humans the most.

The Large Sky Area Multi-Object Fiber Spectroscopic Telescope (LAMOST, also named Guoshoujing telescope) is a special reflecting Schmidt telescope developed by China, which allows both a large aperture (effective aperture of 3.6 m–4.9 m) and a wide field of view ($5^{\circ}$)(Cui et al. 2012). Zhao et al. (2012) reported that LAMOST would enable research in a number of contemporary cutting edge topics in astrophysics, including the formation and evolution of galaxies. From the pilot survey (October 2011) to the eleventh year survey (June 2023), for the low resolution search (LRS) catalog, the LAMOST data released 11 (dr11, "https://www.lamost.org/dr11/v1.0/") has released 11,939,296 spectra, including 11,581,542 stars, 275,302 galaxies, and 82,452 quasars, and so on. The rich spectral survey data is helpful for conducting astrophysical statistical research work. The LAMOST LRS catalog DR11 contains 28,780 stellar population synthesis of galaxies. Together with high redshift quasar spectra, the galaxy data can be used for preliminary cosmological analysis. The preliminary cosmological analysis mainly focuses on the relationship between the galaxy velocity dispersion and the galaxy age, the relationship between the galaxy velocity dispersion and the galaxy metallicity, and the Lyman-$\alpha$ emission lines in spectra of distant quasars. Statistical analysis methods are adopted. In Sect. 2, we introduce the basic information of LAMOST LRS catalogue of stellar population synthesis of galaxies. The preliminary cosmological analysis are performed in Sect. 3, including galaxy age, galaxy velocity dispersion, galaxy metallicity, galaxy redshift, and quasar spectra. In Sect. 4, we summarize our conclusions and perform some discussions.

\section{LAMOST LRS catalog of stellar population synthesis of galaxies}

The LAMOST LRS catalog DR11 has released spectra of 28,780 stellar population synthesis of galaxies. The fluxes of galaxy spectra from LAMOST are calibrated by the photometry data from SDSS (Wang et al. 2018). Based on the data-analysis pipeline published by Westfall et al. (2019), the total integrated fluxes and equivalent widths (EWs) of eight emission lines are calculated. The eight emission lines are $H_{\beta}$, OIII4960, OIII5008, NII6550, $H_{\alpha}$, NII6585, SII6718, and SII6733. Wilkinson et al. (2017) reported a spectral fitting code FIREFLY to derive stellar population properties of stellar systems. Based on the iterative best fitting process controlled by the Bayesian information criterion, FIREFLY fits combinations of single-burst stellar population models to spectroscopic data through a chi-squared minimization fitting. For the stellar population synthesis of galaxies released by the LAMOST LRS catalog DR11, the light-weighted ages and metallicities and the mass-weighted ages and metallicities are obtained by the FIREFLY code. Cappellari (2017) summarized a penalized pixel-fitting (PPXF) method to extract the stellar and gas kinematics, as well as the stellar population of galaxies, via full spectrum fitting. The stellar velocity dispersions of the central galactic regions ($V_{sig}$, Napolitano et al. 2020) and their errors of galaxies from LAMOST LRS catalog DR11 are derived from the PPXF package. Koleva et al. (2008) reported that using different stellar spectral synthesis models to study galaxy age and metallicity can lead to certain systematic biases. Cappelari (2023) used an updated PPXF to study 3200 galaxies with redshift z from 0.6 to 1.0 and obtained stellar velocity dispersions consistent with the results obtained using the Jeans Anisotropic Modelling method. We plan to strictly select galaxy samples and conduct preliminary statistical analysis on these galaxy parameters.

Firstly, we study the galaxy parameters of z, the error of z, the light-weighted age, the mass-weighted age, the light-weighted metallicity, the mass-weighted metallicity, $V_{sig}$, and the errors of $V_{sig}$. Some data processing work are performed. LAMOST marked some unmeasurable parameters as -9999.00. After subtracting these parameters, there are still 28,474 galaxies left. For the redshift z, the galaxies with the error of z less than or equal to 5\% of the value of z are selected. For the stellar velocity dispersion $V_{sig}$, the galaxies with the error of $V_{sig}$ less than or equal to 5\% of $V_{sig}$ are selected. For the galaxy ages, we select the galaxies with the differences between light-weighted ages and mass-weighted ages less than or equal to 15\% of the light-weighted ages, and 15\% of the mass-weighted ages. The light-weighted ages or the mass-weighted ages plus the differences between them are less than 13.8\,Gyr. For the galaxy metallicity, we select the galaxies with the differences between light-weighted metallicities and mass-weighted metallicities less than or equal to 0.1. In addition, there are a few galaxies with $V_{sig}$ greater than 350\,km/s or matallicity less than -0.5. We select the galaxies with $V_{sig}$ less than or equal to 350\,km/s and matallicity greater than or equal to -0.5 to do the research. There are 4,699 galaxies left for statistical analysis. The sharp decrease in data samples is mainly caused by the limitation of galaxy age differences. The 15\% age difference is not very small, but we also need to consider the statistical law of the data samples at the same time. If we adopt the difference in metallicity as less than 0.05, there will be 3259 remaining galaxies. If we take a 10\% age difference, there will be 2997 remaining galaxies. The basic information of the selected galaxy data are shown in Table 1.

\begin{table*}
\caption{The basic information of selected 4,699 galaxies.}
\begin{center}
\begin{tabular}{llllllllllll}
\hline
              &S/N of u filter    &S/N of g filter   &S/N of r filter   &S/N of i filter          &S/N of z filter          &                \\
\hline
main range    &0.5 to 3.0         &10 to 25          &20 to 60          &30 to 90                 &15 to 70                 &                \\
\hline
              &redshift z         &$age_{lw}$[Gyr]   &$age_{mw}$[Gyr]   &$metallicity_{lw}$       &$metallicity_{mw}$       &$V_{sig}$[km/s] \\
\hline
range         &0.03 to 0.17       &0.20 to 12.98     &0.2 to 12.98      &-0.45 to 0.35            &-0.45 to 0.35            &65 to 340       \\  
\hline
\end{tabular}
\end{center}
\end{table*}

Then we performed a preliminary study on the Gunn-Peterson trough and the Lyman-$\alpha$ forest of some high redshift quasar spectra. Finally, we conducted exploratory research on the EW distribution of emission lines of 28,780 galaxies and the diagnostic BPT (Baldwin et al. 1981) diagram for 15,561 galaxies with positive EW values.

\section{The preliminary cosmological analysis}

Thousands of extragalactic systems should be subject to statistical law. Based on the LAMOST LRS catalog of stellar population synthesis of galaxies, we plan to perform a preliminary cosmological analysis.

\subsection{The analysis of stellar velocity dispersion based on the ages of galaxies}

\begin{figure*}
\begin{center}
\includegraphics[width=12cm,angle=0]{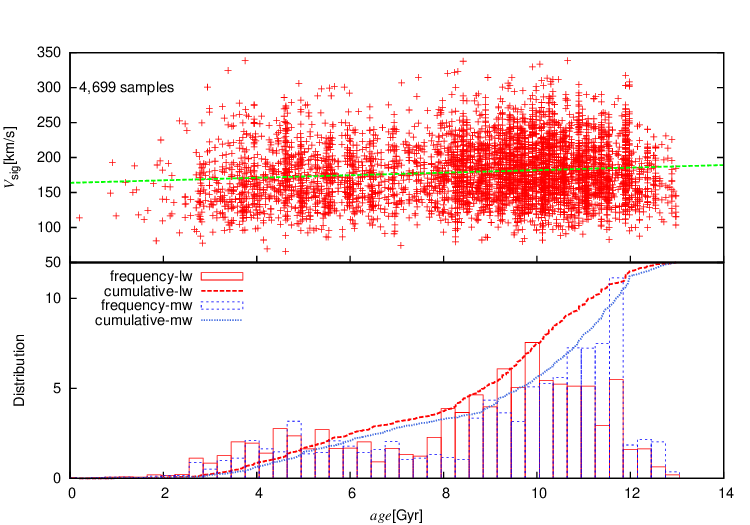}
\end{center}
\caption{The distribution of galaxy age and $V_{sig}$ vs galaxy age diagram for 4,699 galaxies. In the lower panel, the width of each histogram bar is set to 0.3\,Gyr, the cumulative total number of galaxies is set to 12 on the ordinate. The red and blue are for the light-weighted ages and mass-weighted ages respectively. In the upper panel, the green fitting function is $V_{sig}$ = (1.814$\pm$0.235) $\times$ age + (163.850$\pm$2.133) for the lighted-weighted ages with a Pearson correlation coefficient of 0.11. For the mass-weighted ages, it is $V_{sig}$ = (1.806$\pm$0.228) $\times$ age + (162.842$\pm$2.207) with a Pearson correlation coefficient of 0.11.}
\end{figure*}

In Fig. 1, we show a diagram of the distribution of galaxy age and $V_{sig}$ vs galaxy age for 4,699 galaxies. The galaxy ages range from 0.20\,Gyr to 12.98\,Gyr, with some young and some old. The oldest galaxy now is JADES-GS-z13-0 with z = 13.20 and galaxy age of 13.48\,Gyr (Curtis-Lake et al. 2023, Witze 2023). In the lower panel of Fig. 1, we can see that there is a significant proportion of galaxies over 8\,Gyr. In the upper panel, the stellar velocity dispersions of galaxies are fitted by a line function of $V_{sig}$ = (1.814$\pm$0.235) $\times$ age + (163.850$\pm$2.133) with a Pearson correlation coefficient of 0.11. The correlation is weak, which may be due to the small sample size and large age error. According to the positive slope, qualitatively speaking, older galaxies have slightly larger stellar dispersion velocities statistically.

A large number of observational facts and theories of stellar structure and evolution tell us that stars are born in nebulae. Overall, the contraction of nebulae releases gravitational potential energy, which is converted into heat energy. When the temperature is high enough, it will cause hydrogen to ignite, marking the birth of stars. The galaxies are formed in a "bottom up" manner, namely a hierarchical merging theory, as discussed in the introduction. The age of the universe ($t_{0}$) can be calculated using the following equation,
{\small
\begin{equation}
\begin{aligned}
&t_{0}=\frac{1}{H_{0}}\times\\
&{\int_{0}}^{\infty}\frac{1}{(1\!+\!z)(\sqrt{\Omega_{\Lambda}\!+\!\Omega_{k}(1\!+\!z)^{2}\!+\!\Omega_{m}(1\!+\!z)^{3}\!+\!\Omega_{r}(1\!+\!z)^{4}})}dz.\\
\end{aligned}
\end{equation}
}
\noindent In Eq.\,(1), $H_{0}$ is the Hubble constant, $\Omega_{\Lambda}$, $\Omega_{k}$, $\Omega_{m}$, and $\Omega_{r}$ is the dark energy density parameter, the curvature parameter, the matter density parameter, and the radiation density parameter respectively. If we adopt $H_{0}$ = 67.8 $km$ $s^{-1}$ $Mpc^{-1}$ (Ade et al. 2016), $\Omega_{k}$ = 0, $\Omega_{r}$ = 0, $\Omega_{\Lambda}$ = 0.685, and $\Omega_{m}$ = 0.315, the age of the universe is 13.7\,Gyr according to Eq\,(1). If we take the integral in Eq\,(1) from 0.5 to infinity, we can obtain an age of 8.5 Gyr, which is the moment when the universe as space transitioned from being dominated by matter to dominated by dark energy. The dark energy has dominated the universe for 5.2 Gyr and the universe will continue to expand in an accelerated way. The space itself will continue to expand, we infer that the birth rates of galaxies and stars will decrease. After the expansion of space, both the merging efficiency of galaxies and the contraction efficiency of nebulae will decrease. In the lower panel of Fig. 1, we can see that there are significantly more galaxies older than 5.2\,Gyr. Namely, the birth rate of galaxies significantly decreased during the epoch dominated by dark energy. Studying the cosmic star formation history, Madau \& Dickinson (2014) reported that the star formation rate density had a peak at $\sim$3.5\,Gyr after the Big Bang, around z = 1.9. The birth rate of stars decreased during the epoch dominated by dark energy. The statistical results of galaxy age in the lower panel in Fig. 1 are qualitatively consistent with that of SDSS-MaNGA (Mapping Nearby Galaxies at Apache Point Observatory) (Biswas \& Wadadekar 2024).

In the upper panel of Fig. 1, older galaxies have slightly larger stellar dispersion velocities statistically. Both magnetohydrodynamic simulations (IllustrisTNG, Sohn et al. 2024) and observational statistics (ATLAS$^{3D}$, Cappellari et al. 2013) indicate that more massive galaxies have greater stellar dispersion velocities. The stellar dispersion velocity of a galaxy is driven by the total mass of the galaxy. When studying the stellar populations of galaxies in LAMOST LRC DR7 for the first time, Wang et al. (2022) found that the age of galaxies increases with the increase of galaxy mass. Therefore, the relationship presented in the upper panel of Fig. 1 is scientific and reasonable.

\subsection{The analysis of stellar velocity dispersion based on the metallicities}

\begin{figure*}
\begin{center}
\includegraphics[width=12cm,angle=0]{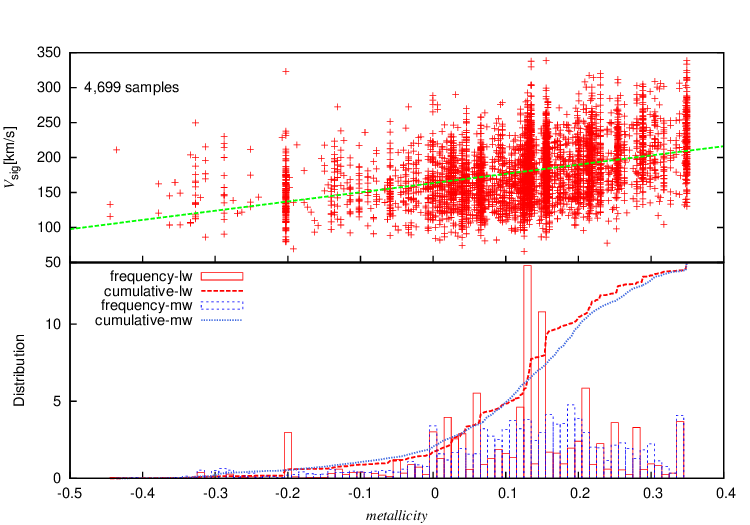}
\end{center}
\caption{The distribution of galaxy metallicity and $V_{sig}$ vs galaxy metallicity diagram for 4,699 galaxies. The width of each histogram bar is set to 0.01. The cumulative total number of galaxies is set to 14 on the ordinate. The green fitting function in the upper panel is $V_{sig}$ = (131.707$\pm$4.539) $\times$ metallicity + (163.523$\pm$0.795) for the light-weighted metallicity with a Pearson correlation coefficient of 0.39. For the mass-weighted metallicity, it is $V_{sig}$ = (111.279$\pm$4.103) $\times$ metallicity + (165.534$\pm$0.775) with a Pearson correlation coefficient of 0.37.}
\end{figure*}

\begin{figure*}
\begin{center}
\includegraphics[width=10cm,angle=0]{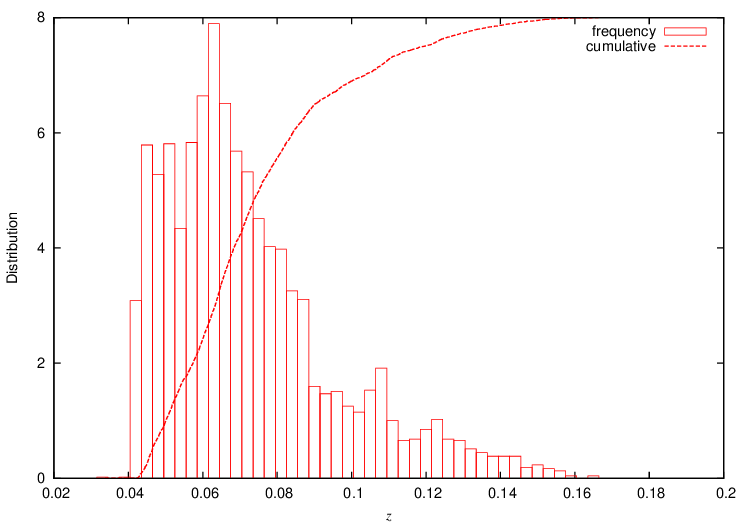}
\end{center}
\caption{The distribution of z for 4,699 galaxies.}
\end{figure*}

In Fig. 2, we show a diagram of the distribution of galaxy metallicity and $V_{sig}$ vs galaxy metallicity for 4,699 galaxies. The galaxy metallicities range from -0.45 to 0.35, with some poor and some rich. Most of the galaxies have metallicities greater than 0.0, even 0.1, in the lower panel, being metal rich galaxies. In the upper panel, the fitting function is $V_{sig}$ = (131.707$\pm$4.539) $\times$ metallicity + (163.523$\pm$0.795) with a Pearson correlation coefficient of 0.39. Statistically, we can see that the larger the metallicity, the greater the $V_{sig}$. This effect is more significant than that in the upper panel of Fig. 1. The relation is caused by the widely known mass-metallicity relation (Kewley \& Ellison 2008, CALIFA galaxies by Garay-Solis et al. 2024), namely more massive galaxies have larger metallicities. In addition, more massive galaxies have larger stellar velocity dispersions, as discussed above. In fact, the statistical relationships in Fig. 1 and 2 are reasonable and consistent with previous work. As time goes on, galaxies undergo processes of element enrichment, resulting in increases in their metallicities. The stellar mass of a galaxy will also increase with time as galaxies are built through merging processes (Kewley \& Ellison 2008).

In Fig. 3, we show a distribution of z for 4,699 galaxies. The value of z ranges basically from 0.04 to 0.16, with most values being less than 0.1. The distant galaxies should be more luminous and massive objects, and there may be a certain selection effect. The analysis of $V_{sig}$ vs z has not been carried out.

\subsection{The cosmological analysis of quasar spectra and galaxy spectra}

\begin{figure*}
\begin{center}
\includegraphics[width=12cm,angle=0]{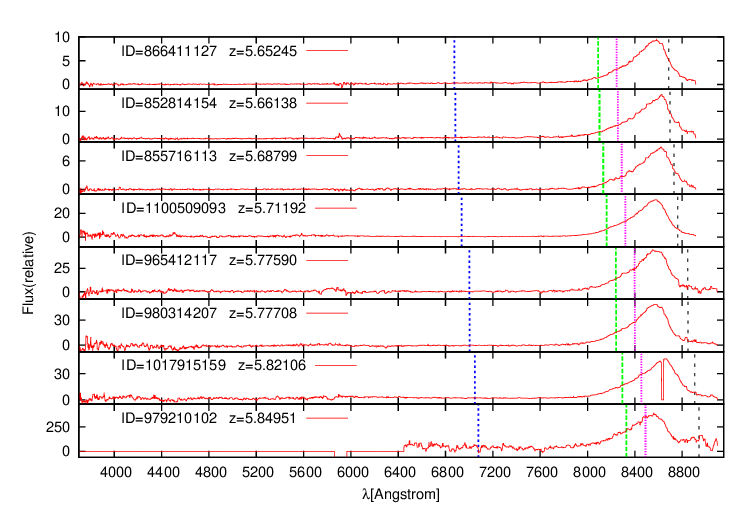}
\end{center}
\caption{Spectra of high redshift quasars from LAMOST. The redshift value is from 5.65 to 5.85. The blue, green, pink, and black dashed lines are the OVI, Lyman-$\alpha$, NV, and OI emission lines, respectively. The spectra are used to show the Gunn-Peterson trough.}
\end{figure*}

\begin{figure*}
\begin{center}
\includegraphics[width=12cm,angle=0]{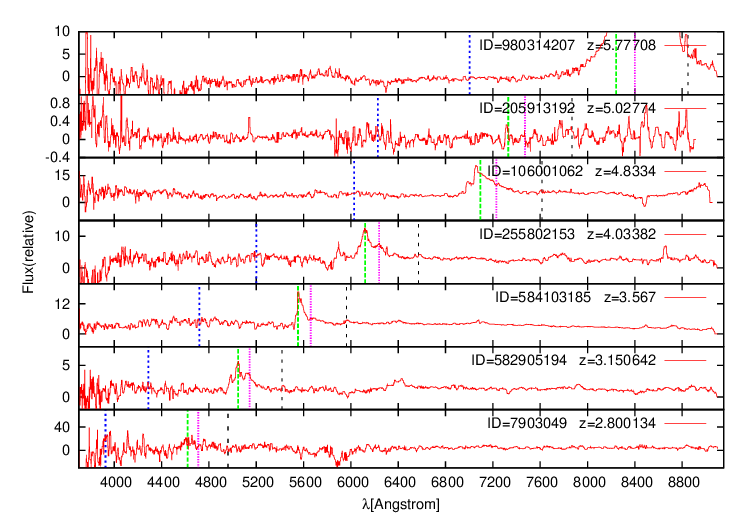}
\end{center}
\caption{Spectra of quasars with redshift value from 2.80 to 5.78. The blue, green, pink, and black dashed lines are the OVI, Lyman-$\alpha$, NV, and OI emission lines, respectively. The spectra are used to show the Lyman-$\alpha$ forest.}
\end{figure*}

\begin{figure*}
\begin{center}
\includegraphics[width=12cm,angle=0]{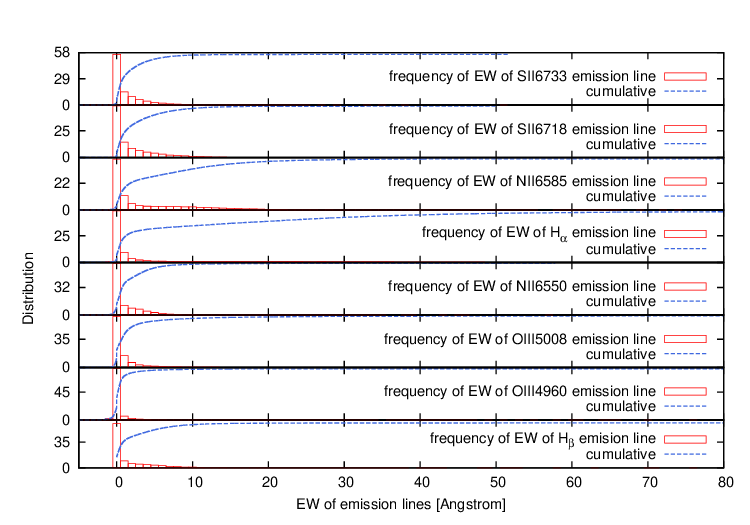}
\end{center}
\caption{The distribution of EW of emission lines of 28,780 galaxies.}
\end{figure*}

\begin{figure*}
\begin{center}
\includegraphics[width=12cm,angle=0]{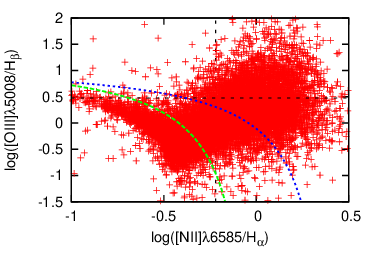}
\end{center}
\caption{The diagnostic BPT (Baldwin et al. 1981) diagram for our 15,561 samples. The blue dashed curve and the green dashed curve shows the demarcation between star-forming galaxies and AGNs by Kewley et al. (2001) and Kauffmann et al. (2003), respectively.}
\end{figure*}

The reionization epoch was observed to start from z = 13.20 and to end at z $\sim$ 6 to 8, as mentioned in the introduction. The Gunn-Peterson trough can be used to study the completion of the reionization. We select quasars with z $>$ 5.6 from LAMOST to make Fig. 4. There are 8 spectra of quasars with z from 5.65 to 5.85 in Fig. 4. The spectral data are 15 pixels smoothed and the effect of Gunn-Peterson trough is obvious, especially for the quasar with ID=1100509093. Through visual inspection, we found that the spectra on the left side of the Lyman-$\alpha$ emission lines (vertical green dashed lines) of the upper 7 quasars were significantly suppressed. It indicates that the universe is in the final stage of the reionization epoch.

In Fig. 5, we show some spectra of quasars with z from 2.80 to 5.78 from LAMOST. The spectra are 15 pixels smoothed. The vertical green dashed line is the Lyman-$\alpha$ emission line, and its left is the Lyman-$\alpha$ forest. For the LRS catalog of LAMOST, the spectra cover the wavelength range of 3700 to 9000\,Å with a resolution of 1800 at 5500\,Å (Stoughton et al. 2002, Abazajian et al. 2003). The Lyman-$\alpha$ forest is obvious, especially for the quasars with ID=255802153 and 584103185. Because the spectral lines on both sides of the vertical green dashed line have a clear contrast for the quasars with ID=255802153 and 584103185. The quasars are very far from the Earth, and their light needs to pass through many extragalactic systems and intergalactic media to reach the Earth. The Lyman-$\alpha$ forest is believed to be mainly caused by neutral hydrogen of extragalactic systems and intergalactic media in the optical pathway.

The LRS catalog quasar spectra released by LAMOST can be used to study the Gunn-Peterson trough and the Lyman-$\alpha$ forest, and thus the end of reionization epoch and the interstellar medium. The studies of the Gunn-Peterson trough and the Lyman-$\alpha$ forest in different directions and different regions can support the uniform and isotropy of the universe on large-scale structure. With the equation of optical depth $\tau$=-ln$\frac{Flux_{obs}}{Flux_{cont}}$ ($Flux_{obs}$ is the observed flux and $Flux_{cont}$ is the continuum flux), we have preliminarily calculated the optical depth of the quasar ID=255802153 (z = 4.03) with wavelengths ranging from 5235 to 5279\,Å. The optical depth follows a Gaussian distribution from 0 to 2, with the lowest end of the Lyman-$\alpha$ absorption line having an optical depth of $\sim$2. The optical depth of the quasar ID=584103185 (z = 3.57), in the wavelength range from 4760 to 4817\,Å, is from 0.0 to 0.8. These results are basically consistent with Eq.\,(12) of Madau (1995) and Eq.\,(5) of Press et al. (1993) for the Lyman-$\alpha$ forest. This is a preliminary research work, and further in-depth spectroscopic research will be carried out in the future.

In Fig. 6, the distribution of EW of emission lines of 28,780 galaxies is displayed. The eight emission lines are $H_{\beta}$, OIII4960, OIII5008, NII6550, $H_{\alpha}$, NII6585, SII6718, and SII6733 respectively, which are from the stellar population synthesis of galaxies released by LAMOST LRS DR11. The proportion of $H_{\beta}$ and $H_{\alpha}$ lines with larger EWs is relatively high, which can be seen in the shape of blue cumulative lines. Obviously, the proportion is significantly larger than that of OIII lines. It indicates that there is indeed abundant neutral hydrogen in the extragalactic systems. The neutral hydrogen can be used to produce the Lyman-$\alpha$ forest. The hydrogen atomic spectroscopy has indeed a wide range of applications in the universe.

We took the positive data (15,561 samples) from Fig. 6 and plotted a diagnostic BPT (Baldwin et al. 1981) diagram, Fig. 7. The blue dashed curve and the green dashed curve shows the demarcation between star-forming galaxies and AGNs by Kewley et al. (2001) and Kauffmann et al. (2003), respectively. There are 9,726 galaxies above the green dashed curve, believed to be AGNs. While, there are 5,835 galaxies below the green dashed curve, believed to be star-forming galaxies. According to the classification method of Kauffmann et al. (2003), there are 2,103 Seyfert galaxies and 4,901 LINERs (low-ionization nuclear emission regions) identified. The emission lines contribute to the exploration and research of galaxies.

\section{Discussion and conclusions}

Cosmology is very interesting. Both the evolution of the universe and that of the galaxies are one of the fundamental issues that we humans are most interested in. The study of high redshift celestial bodies contributes to a more in-depth study of cosmology. The LAMOST LRS catalog DR11 v1.0 has released spectra of 28,780 stellar population synthesis of galaxies and 82,452 quasars. We have conducted exploratory research on these extragalactic systems and some high redshift quasars from a cosmological perspective. The 28,780 stellar population synthesis of galaxies contain data of redshift, the error of redshift, light-weighted age, mass-weighted age, light-weighted metallicity, mass-weighted metallicity, stellar velocity dispersion, and the error of stellar velocity dispersion. We select the data with small errors, especially the differences between light-weighted ages and mass-weighted ages less than or equal to 15\% of the light-weighted ages, and 15\% of the mass-weighted ages. There are a total of 4,699 galaxy data being left and used for statistical analysis.

In Fig. 1 and 2, the stellar velocity dispersion is directly proportional to the galaxy age and the galaxy metallicity, respectively. The small Pearson correlation coefficients may be due to the insufficient sample size or slightly larger age errors. Although the distributions of $V_{sig}$ for galaxies have large dispersions, the statistical law of the fitting functions are determined. We attempted to relax the selecting conditions and increase the number of data samples, and then obtained the same conclusions. This is due to the fact that more massive galaxies have greater stellar velocity dispersions (Sohn et al. 2024, Cappellari et al. 2013), relatively older ages, and greater metallicities (Wang et al. 2022). This is reasonable, as galaxies undergo a chemical enrichment process with increasing age, leading to an increase in metallicity.  The stellar mass of a galaxy will increase with time as galaxies are built through merging processes (Kewley \& Ellison 2008). Our statistical results are indirectly consistent with the hierarchical merging theory. Perhaps due to the selection effect of samples being too bright and too close neighbors, no obvious morphological signatures of galaxy mergers were observed. In the future, we will use LAMOST together with complementary surveys, such as deeper photometry or IFS data to tighten the correlation analyses or better measure galaxy masses.

In addition, we use the spectra of quasars with high value of z to study the Gunn-Peterson trough and the Lyman-$\alpha$ forest. The Gunn-Peterson trough provides strong evidence for the end state of the cosmic reionization (Fan et al. 2006). The Lyman-$\alpha$ forest indicates the abundance of neutral hydrogen of extragalactic systems and intergalactic media in the optical pathway. The equivalent widths of emission lines reflect the abundance of neutral hydrogen in the extragalactic systems. According to the diagnostic BPT diagram (Fig. 7), there are 9,726 AGNs and 5,835 star-forming galaxies in our samples. The high redshift celestial spectra and the identified emission lines released by LAMOST can be used for cosmological research.

\section*{Acknowledgments}

Guoshoujing Telescope (the Large Sky Area Multi-Object Fiber Spectroscopic Telescope LAMOST) is a National Major Scientific Project built by the Chinese Academy of Sciences. Funding for the project has been provided by the National Development and Reform Commission. LAMOST is operated and managed by the National Astronomical Observatories, Chinese Academy of Sciences. The work is supported by the Yunnan Provincial Department of Education Science Research Fund Project (No. 2024J0964) and the International Centre of Supernovae, Yunnan Key Laboratory (No. 202302AN36000101).

%\section*{Data availability}
%The data underlying this article are available in the article and in its online supplementary material.



%%%%%%%%%%%%%%%%%%%%%%%%%%%%%%%%%%%%%%%%%%%%%%%%%%

% Don't change these lines
%\bsp	% typesetting comment
%\label{lastpage}
\end{document}